# CIV Absorption From Galaxies in the Process of Formation

Martin G. Haehnelt[1], Matthias Steinmetz[1] & Michael Rauch[2,3]

mhaehnelt@mpa-garching.mpg.de, msteinmetz@mpa-garching.mpg.de, mr@astro.caltech.edu

## Abstract

We investigate the heavy element QSO absorption systems caused by gas condensations at high redshift which evolve into galaxies with circular velocity of 100 to 200 km s$^{-1}$ at the present epoch. Artificial QSO spectra were generated for a variety of lines-of-sight through regions of the universe simulated with a hydrodynamics code. The CIV and HI absorption features in these spectra closely resemble observed CIV and HI absorption systems over a wide range in column density if [C/H] $\sim -3$ to $-2$ is assumed. CIV absorption complexes with multiple-component structure and velocity spreads up to $\sim 600$ km s$^{-1}$ are found. The broadest systems are caused by lines-of-sight passing through groups of protogalactic clumps with individual velocity dispersions of less than 150 km s$^{-1}$ aligned along filamentary structures. The largest velocity spreads are kinematic and due to the Hubble expansion between different clumps. Typical clumps have a radius of about $5 - 10$ kpc, a baryonic mass of $\sim 10^9$ M$_\odot$ and are surrounded by hot gas atmospheres with radius of about 30 kpc. The temperature of most of the gas does not take the photoionization equilibrium value. This invalidates density and size estimates derived from thermal equilibrium models. Consequences for metal abundance determinations are briefly discussed. We predict occasional exceptionally large ratios of CIV to HI column density (up to a third) for lines-of-sight passing through compact halos of hot gas with temperature close to $3 \times 10^5$ K. Our model may be able to explain both high-ionization multi-component heavy-element absorbers and damped Ly$\alpha$ systems as groups of small protogalactic clumps.

[1]Max-Planck-Institut für Astrophysik, Karl-Schwarzschild-Straße 1, Garching, Germany

[2]Astronomy Dept., 105-24 California Institute of Technology, Pasadena 91125, USA

[3]Hubble Fellow

astro-ph/9512118  18 Dec 1995

# 1 Introduction

Heavy element absorption as seen in the spectra of distant quasi-stellar objects (QSOs) is commonly thought to be produced by metal-contaminated gas associated with galaxies. Galactic halo clouds, disks, HII regions, winds, tidal tails, and satellites have all been invoked as possible sites producing absorption. A connection of metal absorption systems with observable galaxies is strongly indicated by successful searches for galaxies close to the absorption redshift (Bergeron 1986; Bergeron & Boissé 1991; Lanzetta et al. 1995; Steidel 1995). These surveys for galaxies related to low-ionization systems at low and intermediate redshifts indicate that these objects may be associated with normal (i.e. redshift zero) galaxies, and Wolfe (1995) and collaborators have argued that the very low ionization, damped Ly$\alpha$ systems are progenitors of present-day disk galaxies.

However, such a close one-to-one correspondence between galaxy and absorption system need not be valid for higher-ionization (CIV) systems at higher redshifts (York 1988). Several features like the extended component structure of CIV selected absorption systems are equally difficult to understand in terms of a stationary cloudlet-in-halo model as in terms of absorption along lines-of-sight (LOS) through known types of local galaxy clusters (Pettini et al. 1983; Morris et al. 1986). Instead, going back in time we should ultimately see the small-scale gas condensations from which instabilities and hierarchical clustering have build up present-day galaxies. Observational parameters of QSO absorbers (Sargent, Boksenberg & Steidel 1988; Steidel 1990) are at least consistent with heavy-element systems arising in dynamically evolving clouds which form early on during the collapse of galaxies (see also Lake 1988).

The use of numerical simulations to model the large-scale distribution of neutral hydrogen has recently lead to a major advance in the understanding of Ly$\alpha$ absorption systems (Cen et al. 1994; Petitjean, Mücket & Kates 1995; Hernquist et al. 1995). Here we investigate smaller scales in order to understand the nature of heavy element absorption systems by tracing the history of present-day galaxy halos back in time. For this purpose we have used a set of Smoothed-Particle-Hydrodynamics (SPH) simulations to model the detailed evolution of regions with comoving radius of about 3 Mpc which by the present epoch contain one to three galaxies with halo circular velocities between 100 and 200 km s$^{-1}$. We focus on the CIV absorption properties of these regions. Recent high signal-to-noise QSO spectroscopic studies by Tytler et al. (1995) and Cowie et al. (1995) find CIV absorption down to rather smaller HI column densities than previous work, indicating that all baryonic matter at the relevant redshift could be contaminated with heavy elements to some degree. For simplicity we assume in the present paper that all gas is homogeneously contaminated with a small fraction of metals.

In section 2 we describe the technical details of the SPH simulations and of our determination of the ionization state of carbon. Section 3 illustrates the characteristic features of CIV absorption in our simulations and in section 4 we discuss some of the implications.

## 2 The spatial distribution of neutral hydrogen and CIV

### 2.1 Simulating the gas distribution

The simulations were performed using GRAPESPH (Steinmetz 1996). The cosmological background model is a $\Omega = 1$, $H_0 = 50 \, \mathrm{km\,s^{-1}\,Mpc^{-1}}$ cold dark matter (CDM) cosmogony with a normalization of $\sigma_8 = 0.63$. The baryon fraction is $\Omega_b = 0.05$. The simulations are similar to the galaxy formation simulations performed by Navarro, Frenk & White (1995), but the mass resolution (gas particle masses of $5 \times 10^6 \, \mathrm{M}_\odot$ to $2 \times 10^7 \, \mathrm{M}_\odot$) is superior by a factor of 6. The gravitational softening is 1.25–2.5 kpc for the gas and a factor of two larger for the dark matter. The simulations cover a high resolution sphere with a radius of about 2–4 Mpc. The tidal field due to surrounding matter was taken into account up to a radius of 15 Mpc using the multi-mass technique described in Navarro et al. (1995). All systems which are analyzed below lie in the inner 1.5 Mpc of the high resolution region, i.e. the dynamics is not contaminated by the more massive outer particles. At redshift zero, the high-resolution region contains between one and three disk-like galaxies each consisting of more than 1000 particles. The circular velocities of the dark matter halos lie between 100 and 200 $\mathrm{km\,s^{-1}}$. Pronounced disk-like features do not develop before a redshift of 1.5, while most of the analysis presented below is done at a redshift of 3. For the UV background we assume the redshift dependence given by Vedel, Hellsten & Sommer–Larsen (1994),

$$J(\nu) = 10^{-21} \left[ 1 + \left( \frac{5}{1+z} \right)^4 \right]^{-1} \left( \frac{\nu}{\nu_H} \right)^{-\alpha} \mathrm{erg\,s^{-1}\,cm^{-2}\,Hz^{-1}\,str^{-1}}, \quad (1)$$

where $\nu_H$ is the frequency corresponding to the ionization energy of hydrogen (13.6 eV). For $\alpha$ a value of 1.5 was usually assumed. This corresponds to a typical quasar-like spectrum and is consistent with measurements of the helium Gunn-Peterson decrement (Tytler et al. 1995). The background was switched on at the beginning of the simulation ($z = 21$) and the initial gas temperature was assumed to take the value, where photoionization heating balances the cooling due to collisional processes and to Compton scattering ($\approx 3 \times 10^4$ K).

### 2.2 The ionization state of the gas

We have used the photoionization code CLOUDY to calculate the ionization state of carbon for the temperature, density, and UV background of each SPH particle in the simulation (Ferland 1993). CLOUDY was used in its "Ly$\alpha$ absorber mode". An infinite slab of gas of low metallicity, optically thin to ionizing radiation and illuminated from both sides by a homogeneous UV field, is then assumed. The solid curves in Figure 1a show the column density ratio of CIV and neutral hydrogen for [C/H] = −2, a UV background given by equation (1) with $\alpha = 1.5$ and $z = 3$, and a fixed temperature as indicated on the plot. For low densities and small temperatures the ionization state is completely determined by the ionization parameter ($\propto J/n$, where $n$ is the gas density) and the curves shift linearly to the right with increasing ionizing flux. For the dashed curve the temperature was assumed to be determined by photoionization equilibrium. Figure 1b shows a typical scatter plot of density versus temperature of the SPH particles in our simulations. The solid curve is the equilibrium temperature. There is a large scatter towards higher temperatures around the peak in the equilibrium curve which is caused by compressional heating because cooling times are comparable to or larger than typical dynamical times. Also seen is a modest deviation from the equilibrium curve towards smaller temperatures due to adiabatic cooling at the lowest densities. The dotted curves in Figure 1a are for an extremely soft spectrum with $\alpha = 5$. In this case it is difficult to ionize carbon beyond CIV. The CIV/HI column density ratio therefore increases strongly towards lower densities.

## 3 Absorption spectra

### 3.1 General appearance of absorption features

In Figure 2 we show typical absorption systems for a range of HI column densities. The artificial spectra were produced with a signal-to-noise of 50, an instrumental broadening of $8\,\mathrm{km\,s^{-1}}$, and a bin size of $2\,\mathrm{km\,s^{-1}}$ to match the spectral resolution obtainable with the HIRES spectrograph of the Keck telescope. HI and CIV are overplotted in velocity space. For clarity we generally plot only one line of the CIV $\lambda\lambda 1548/1550$ doublet. Tests with simulations of lower resolution showed that we still undersample the numerical resolution. All the features seen in the plots are real and are numerically well resolved.

At column densities of about $10^{14}\,\mathrm{cm^{-2}}$, the CIV absorption traces the HI absorption well and is just marginally detectable. For increasing HI column densities more and more substructure is seen in CIV until a complex multi-component structure of narrow CIV absorption lines is evident at the highest column densities (see also Figure 3). The spectra show a convincing resemblance both to the weak CIV absorption systems observed e.g. by Cowie et al. (1995) and others and the stronger variants seen in low-resolution CIV surveys (e.g. Sargent, Boksenberg & Steidel 1988). We further investigated the effect of varying the strength and spectrum of the UV background. Although clear differences can be seen for individual absorbers, the changes mainly affect the total column density in HI and CIV; the ratio between the two remains roughly the same. For fixed column density the overall appearance does not vary significantly.

In Figure 3 we show the HI and CIV absorption for a $5\times 5$ mosaic of LOS around a filamentary structure at $z = 3$. The proper distance between the panels is 10 kpc. The LOS at the center directly penetrates about 5 to 7 different clumps (see also Figure 4b) producing the kind of prominent clusters of CIV lines discussed above. The clustering pattern varies considerably over scales of 5 kpc. In the corner panels, at a distance of about 30 kpc, the CIV absorption has almost dropped below the detection threshold.

## 3.2 Velocity structure

In Figures 4a and 4b we present two different kinds of structures which we found to be responsible for prominent clusters of narrow CIV absorption lines. The four upper panels show the absorption, the density, the CIV and HI weighted temperatures, and the velocity profile along the LOS. The bottom panels contain CIV density contour maps. The contour plot and the density profile of Figure 4a shows that the LOS penetrates a single clump. The temperature is clearly enhanced above the ionization equilibrium temperature and the complicated temperature profile is responsible for the cluster of CIV lines. The object is obviously not dynamically relaxed. Figure 4b shows a LOS along a filamentary structure. The alignment of 5 to 7 well separated clumps none of which has a circular velocity larger than about $100\,\mathrm{km\,s^{-1}}$ is responsible for the prominent CIV cluster which extends over $500\,\mathrm{km\,s^{-1}}$. The broadest and most complex absorption features found in our simulations are all produced by such an alignment of dense clumps along filaments.

## 3.3 Abundance determinations

Determining metal abundances from QSO absorption lines requires at least one pair of ionic species occupying the same spatial region and experiencing the same physical conditions. For low-column-density systems, however, there is often only the CIV/H ratio available, so the abundance determination on the basis of this value involves considerable uncertainties because of the difference in ionization potentials and the poorly constrained physical conditions. As shown in Figure 1a, for the equilibrium case with temperatures of about $10^4$ K there is a broad peak in the CIV/HI column-density ratio, $\log(\mathrm{CIV/HI}) \sim [\mathrm{C/H}]$, at a density of $10^{-4}\,J_{-21}\,\mathrm{cm}^{-3}$ which extends over two to three orders of magnitude and where $J_{-21}$ is the ionizing flux in units of $10^{-21}\,\mathrm{erg\,cm^{-2}\,s^{-1}\,Hz^{-1}\,sr^{-1}}$. Cowie et al. (1995) therefore infer $[\mathrm{C/H}] = -3$ to $-2$ from their measured CIV/HI column density ratios of about $10^{-3}$ to $10^{-2}$. For our simulations we find that this assumption is often valid. Figure 4c shows a typical low-column density system in detail. The input metallicity can be recovered quite well from such an analysis. However, as shown in Figure 4d, for some LOS this reasoning can fail by a large margin. Despite the fact that $[\mathrm{C/H}] = -2$, the CIV column density here is only a factor eight smaller than the HI column density. These rather rare LOS pass through the hot atmospheres of dense clumps with gas temperatures close to $3 \times 10^5$ K where the CIV/HI ratio reaches a maximum for collisionally ionized gas (see Figure 1a).

## 4 Discussion and Conclusions

We have demonstrated that LOS through regions of ongoing galaxy formation are capable of producing CIV absorption complexes with kinematic structures and column densities very similar to the ones actually observed. The individual CIV components are related to protogalactic condensations (PGC) with baryonic masses of about $10^9\,\mathrm{M}_\odot$ and radii of about $5-10$ kpc.

The PGCs are in a variety of dynamical states, including non-virialized motions due to the merging of several clumps. There seems to be a trend towards more structure in the CIV absorption pattern when the gas distribution along the LOS is less dynamically relaxed. Velocity dispersions in the CIV absorption pattern of a few hundred km/s are often due to small-scale motions of bound objects, while larger velocities usually arise from the large-scale expansion of the universe and the fortuitous orientation of the LOS along an expanding filament. This interpretation avoids the difficulties associated with accounting for systems with velocity extent in excess of $500\,\mathrm{km\,s^{-1}}$ in terms of galaxy clusters or clouds orbiting in massive halos at these high redshifts (see also Pettini et al. 1983 and Morris et al. 1986).

If we choose a LOS through the very center of a PGC, the HI column density easily exceeds $10^{20}\,\mathrm{cm^{-2}}$. Thus our findings may also be relevant for an understanding of damped Ly$\alpha$ systems. The observed component structure of higher ions associated with these systems is very similar to those of non-damped CIV systems. The main difference between CIV-'only' and damped Ly$\alpha$ systems could thus be one of impact parameter and the relatively smaller cross-section for high column densities. This implies a possible unified picture for highly ionized absorbers with multi-component structure and damped Ly$\alpha$ systems. Despite their different column densities both might be caused by groups of PGCs. The much rarer damped systems would then simply be cases with one PGC occurring close enough to our LOS to show damped Ly$\alpha$ absorption. Observational evidence for damped Ly$\alpha$ systems as small as the PGCs discussed here comes from the work of Møller & Warren (1995;Warren & Møller 1995). Theoretical work by Kauffmann (1995) also predicts the objects responsible for damped Ly$\alpha$ alpha systems to be rather smaller than expected if they were large protogalactic disks (Wolfe 1995).

The complex component structures arises from density and temperature profiles which vary strongly along the LOS. The assumption that the temperature takes the photoionization equilibrium value, often made in photoionization models, will fail in such cases (see Miralda-Escudé & Rees 1993 for a discussion of non-equilibrium effects). Using the equilibrium temperature as an indicator of the ionization parameter can give grossly incorrect gas densities due to the very strong dependence of the ionization parameter on the temperature. Furthermore, even for a correctly determined density ( e.g. using the observed CII/CIV column-density ratio), the fraction of neutral hydrogen is systematically overestimated. This leads to an estimate of the cloud size along the LOS which can be easily smaller by a factor of a hundred than the correct value. This probably explains the very small CIV cloud sizes of $\sim$ 100 pc inferred by Cowie et al. (1995) based on equilibrium photoionization models of Donahue & Shull (1991). Standard abundance determinations also generally assume the gas to be close to the photoionization equilibrium temperature. In most cases we expect this to give acceptable abundances. Nevertheless, because of the frequent deviations from the photoionization equilibrium temperature metal abundance determinations on the basis of CIV/HI column density ratios alone should be considered with some caution.

Although very idealized, a homogeneous metal contamination of the intergalactic medium with [C/H] $\sim -2$ leads to an astonishingly good representation of the observed CIV absorption. The appearance of the absorption line complexes is determined mainly by the temperature and ionization structure of the density maxima and by the large-scale dynamics of the PGC groups. Thus we expect that our results will not change much, if the metals are not homogeneously distributed as might be the case if they are produced in small globular-cluster-like objects (Fall & Rees 1985). Radial metallicity gradients and a reduction of [C/H] outside dense clumps would probably have only minor effects. However, an enhancement above the level [C/H] = $-2$ to $-3$ in dense regions ($10^{-5} - 10^{-3}(0.1/J_{-21})$ cm$^{-3}$) at redshift $z = 3$ would lead to overproduction of strong CIV absorbers.

We conclude that the gas physics and the mass scales included in our simulations of regions of ongoing galaxy formation are sufficient to recover the basic properties of observed CIV absorption systems. A detailed statistical analysis of such simulations and an extension to other ionic species is in progress and should give constraints on the history and spatial distribution of star formation in the progenitors of galaxy halos.

## 5 Acknowledgments

We thank Martin Rees and Simon White for very helpful comments and Gary Ferland for making CLOUDY available to us. MR acknowledges instructive conversations with Bob Carswell, Len Cowie, Esther Hu, Palle Møller, Ray Weymann and Wal Sargent, and is grateful to NASA for support through grant HF-01075.01-94A from the Space Telescope Science Institute, which is operated by the Association of Universities for Research in Astronomy, Inc., under NASA contract NAS5-26555. Support by NATO grant CRG 950752 is also gratefully acknowledged. MGH thanks Wal Sargent and Caltech for hospitality.

**Figure captions**

Figure 1: The solid curves in the left panel show the column density ratio CIV/HI as function of density for [C/H] = $-2$, a UV background with $J_{-21} = 0.3$ and spectral index $\alpha = 1.5$ beyond the Lyman edge. Calculations were done with CLOUDY for fixed temperatures as indicated in the plot. For the dashed curve the temperature is determined by photoionization equilibrium. The dotted curves are for an extremely soft spectrum with $\alpha = 5$. The right panel shows a typical temperature/density distribution for the gas particles in our SPH simulations. The solid curves show the corresponding relation assuming photoionization equilibrium.

Figure 2: Typical CIV absorption systems are shown for a range of HI column densities. The spectra have signal-to-noise of 50, an instrumental broadening of $8\,\mathrm{km\,s^{-1}}$, and a bin size of $2\,\mathrm{km\,s^{-1}}$. A flat continuum is assumed. Only one line of the CIV doublet is shown. HI and CIV absorption are shifted relative to each other by 0.5. The log of the HI column densities is indicated on the plot.

Figure 3: CIV and HI absorption for a mosaic of lines-of-sight around a filamentary structure at $z = 3$. HI and CIV absorption are shifted relative to each other by 0.5. Only one line of the CIV doublet is shown. The proper distance between neighbouring panels is 10 kpc.

Figure 4: The top panels show CIV and HI absorption along selected lines-of-sight at $z = 3$. The signal-to-noise is 50, the instrumental broadening $8\,\mathrm{km\,s^{-1}}$, and the bin size $2\,\mathrm{km\,s^{-1}}$. Only one line of the CIV doublet is shown. The next three panels show from top to bottom the corresponding total gas density, HI (solid curve) and CIV (dashed curve) weighted temperatures and the peculiar velocity profile. The bottom panels show contour plots of the CIV density in a plane containing the LOS. The spacing of the contours is 2 dex.

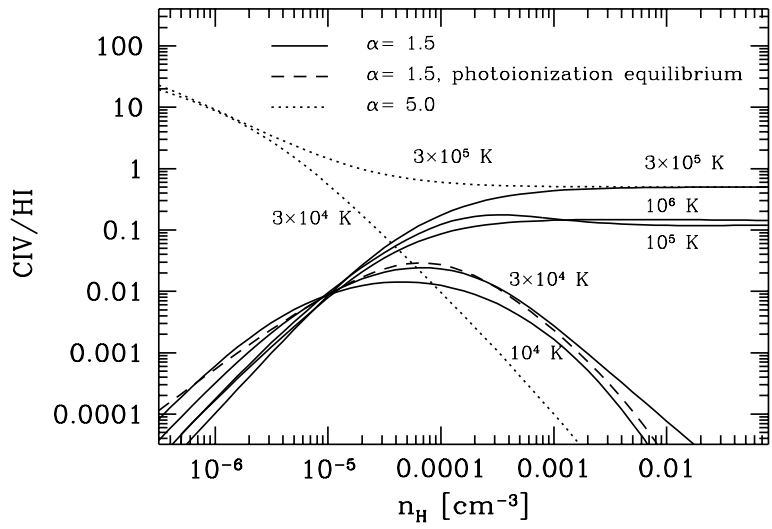

Figure 1a

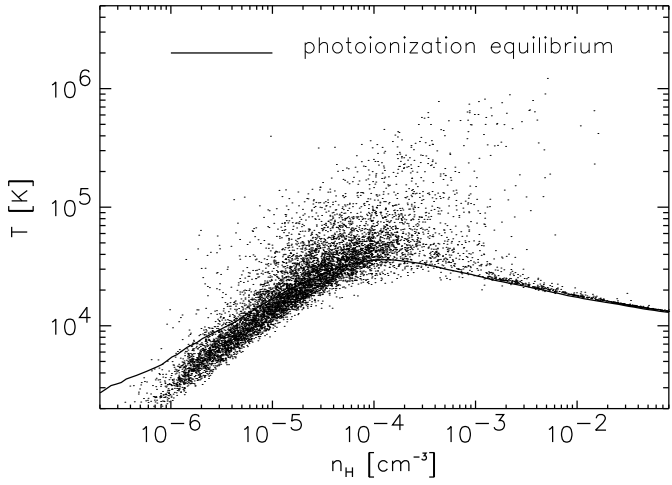

Figure 1b

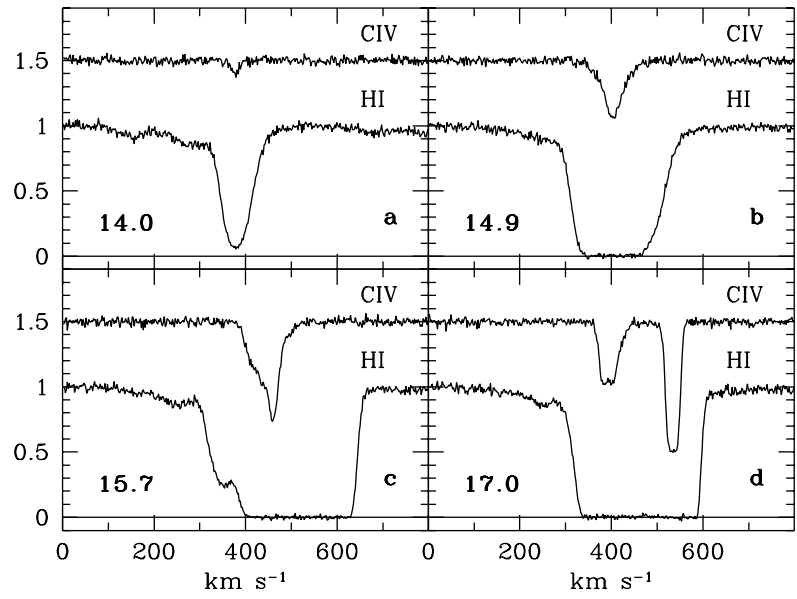

Figure 2

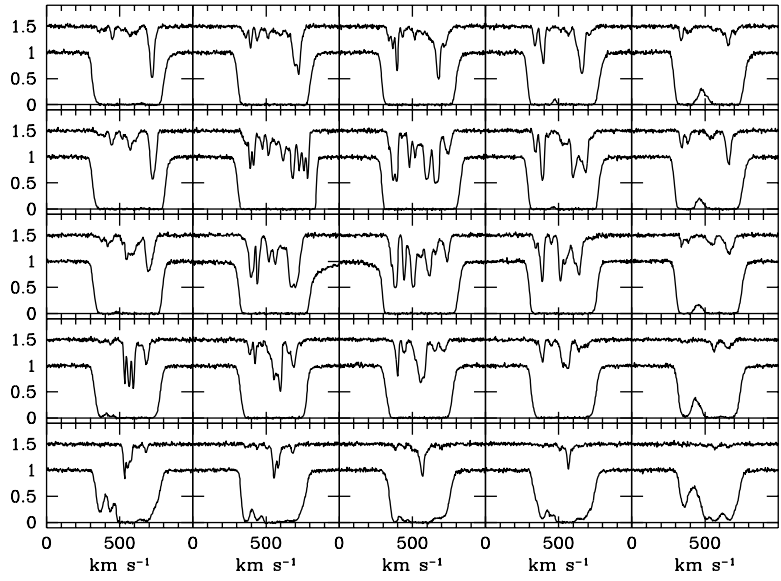

Figure 3

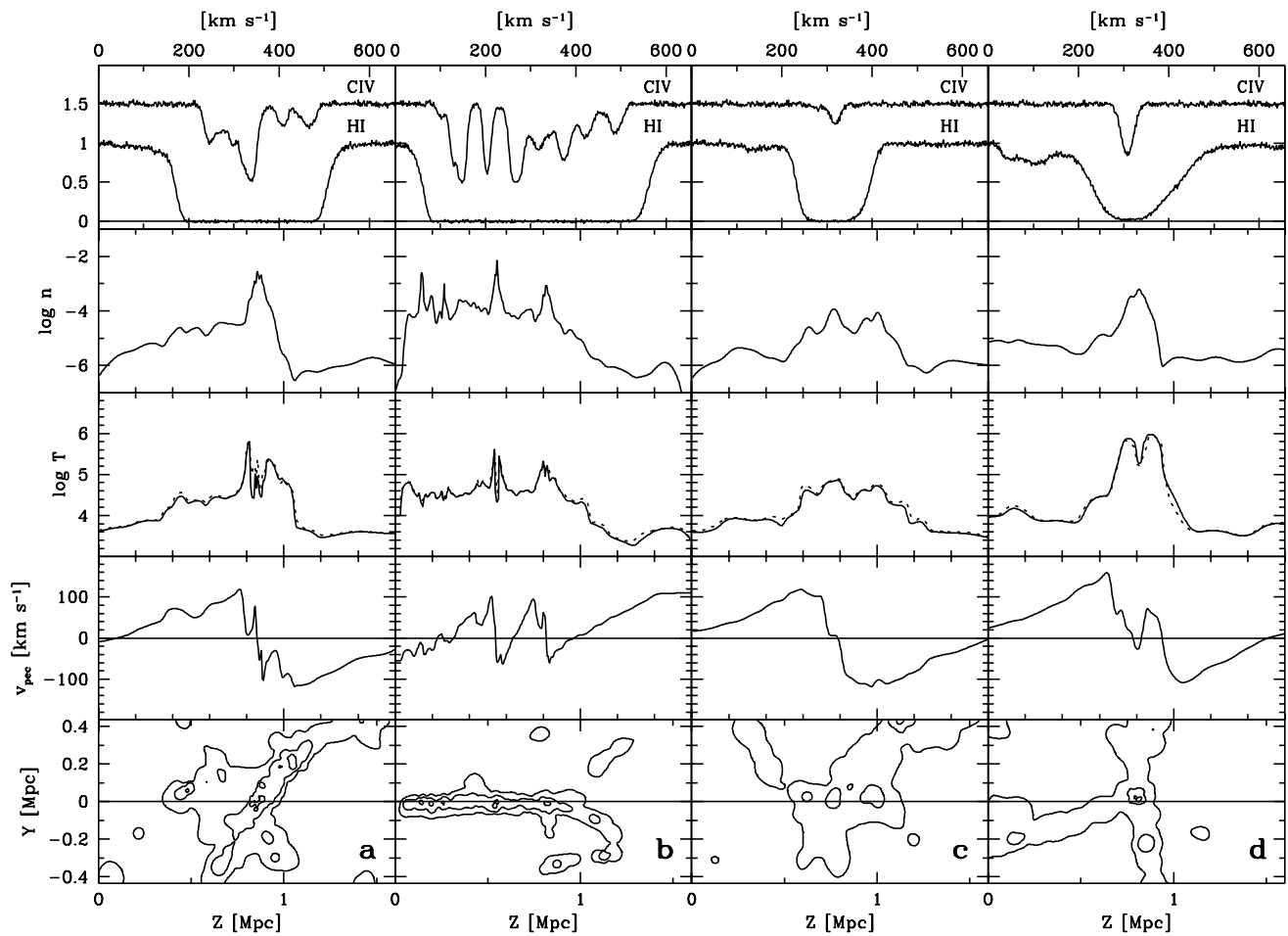

Figure 4